\documentclass[sigconf]{acmart}

\AtBeginDocument{%
  \providecommand\BibTeX{{%
    \normalfont B\kern-0.5em{\scshape i\kern-0.25em b}\kern-0.8em\TeX}}}

\setcopyright{acmcopyright}
\copyrightyear{2018}
\acmYear{2018}
\acmDOI{10.1145/1122445.1122456}

\acmConference[Woodstock '18]{Woodstock '18: ACM Symposium on Neural
  Gaze Detection}{June 03--05, 2018}{Woodstock, NY}
\acmBooktitle{Woodstock '18: ACM Symposium on Neural Gaze Detection,
  June 03--05, 2018, Woodstock, NY}
\acmPrice{15.00}
\acmISBN{978-1-4503-XXXX-X/18/06}

\renewcommand\footnotetextcopyrightpermission[1]{} 
\usepackage{layouts}
\usepackage{titlesec}

\usepackage{fancyhdr}
\usepackage[normalem]{ulem}

\usepackage{algorithmic}
\usepackage{graphicx}
\usepackage{textcomp}
\usepackage{xcolor}

\usepackage{xurl}
\usepackage{etoolbox}

\usepackage{titlesec}
\usepackage{lipsum}
\usepackage{layouts}

\usepackage[utf8]{inputenc}

\usepackage{float}
\usepackage{booktabs} 
\usepackage{multirow}

\usepackage{caption}
\usepackage{subcaption}

\usepackage{makecell}

\usepackage{tabularx,ragged2e,booktabs,caption}

\usepackage{mathtools}

\usepackage{balance}
\usepackage{listings}
\usepackage{color}
\definecolor{dkgreen}{rgb}{0,0.6,0}
\definecolor{gray}{rgb}{0.5,0.5,0.5}
\definecolor{mauve}{rgb}{0.58,0,0.82}
\definecolor{bg}{rgb}{0.9,0.9,0.9}
\definecolor{amber}{rgb}{1.0, 0.75, 0.0}

\PassOptionsToPackage{hyphens}{url}\usepackage{hyperref}

\newcommand{\framework}{\mbox{GNN-DSE}}
\PassOptionsToPackage{hyphens}{url}\usepackage{hyperref}

\usepackage{pifont}

\usepackage{enumitem}
\setlist[itemize,1]{itemsep=0.5pt,partopsep=0pt,parsep=\parskip, topsep=2pt, leftmargin=10pt,}
\setlist[enumerate,1]{itemsep=0.5pt,partopsep=0pt,parsep=\parskip, topsep=2pt, leftmargin=10pt,}

\usepackage[colorinlistoftodos]{todonotes}

\newcommand{\arxiv}[1]{{\color{black}{#1}}}

\usepackage{xspace}
\newcommand{\tconv}{\textsc{TransformerConv}\xspace}
\newcommand{\model}{\textsc{GNN-DSE}\xspace}

\usepackage{bm}

\begin{document}
\title{Enabling Automated FPGA Accelerator Optimization Using Graph Neural Networks}

\author{Atefeh Sohrabizadeh, 
        Yunsheng Bai, Yizhou Sun, and Jason Cong}
\affiliation{%
  \institution{Computer Science Department, University of California - Los Angeles, USA}
  \country{Los Angeles, CA, USA}
}
\email{{atefehsz, yba, yzsun, cong}@cs.ucla.edu}

\pagenumbering{arabic}

\thispagestyle{firstpage}
\pagestyle{plain}


\begin{abstract}
High-level synthesis (HLS) has freed the computer architects from developing their designs in a very low-level language and needing to exactly specify how the data should be transferred in register-level. With the help of HLS, the hardware designers must describe only a high-level behavioral flow of the design. Despite this, it still can take weeks to develop a high-performance architecture mainly because there are many design choices at a higher level that requires more time to explore. It also takes several minutes to hours to get feedback from the HLS tool on the quality of each design candidate. In this paper, we propose to solve this problem by modeling the HLS tool with a graph neural network (GNN) that is trained to be used for a wide range of applications. The experimental results demonstrate that by employing the GNN-based model, we are able to estimate the quality of design in milliseconds with high accuracy which can help us search through the solution space very quickly.
\end{abstract}
\maketitle


\section{Introduction} \label{sec:intro}

The demand for scalable, high-performance computing is increasing rapidly. However, due to the breakdown of Dennard's scaling~\cite{dennard74}, we no longer can address it by scaling the clock frequency. This has led to a growing interest into domain specific computing and exploration into using accelerators such as field-programmable gate arrays (FPGAs) to reduce power consumption while achieving a high performance~\cite{catapult, amazon-f1}. On the downside, the FPGAs are more difficult to program compared to CPUs and GPUs. It used to be case that one only could program an FPGA by writing very low level codes that described the transition of data in register-transfer level (RTL). In the past decades, high-level synthesis (HLS)~\cite{cong11,zhang2008autopilot} was introduced to simplify the programming by raising the abstraction level in FPGA design. With HLS, the designer only needs to describe a high-level behavioral description of the design. As a result, HLS has been embraced by both academia and industry~\cite{hls4ml, lai2019heterocl, sohrabizadeh2020end}. Currently, both FPGA vendors offer their commercial HLS products---Xilinx Vitis~\cite{vitis-platform} and Intel FPGA SDK for OpenCL~\cite{intel-sdk}. 

The HLS tools let the designers optimize their microarchitecture quickly by inserting a few synthesis directives in the form of pragmas. This feature can potentially help to decrease the turn-around times and shorten the code development cycle. However, because the HLS tools look for the right combination of pragmas for deriving an efficient hardware architecture, not every HLS design has a good quality of results (QoR)~\cite{sohrabizadeh2020autodse}. Thus, one often has to explore many design choices for each new application which can negatively impact the design turn-around time.

To speed up design optimization, a new line of research has been created with the focus on automating the microarchitecture optimizations (e.g.~\cite{sohrabizadeh2020autodse, s2fa, kwon2020transfer}). By developing an automated design space exploration (DSE) framework, not only will the hardware designers be free of the design improvement iterations, but other programmers with no knowledge of hardware can also try customized computing. This can in turn result in the growth of the FPGA community and its technology improvement. However, the HLS-based DSE pose the following challenges:

\begin{itemize}
    \item \textbf{The long synthesis time of existing commercial HLS tools:}
Using the vendor HLS tools directly for DSE results in a long evaluation time (minutes to hours) for each design candidate and forces us to explore a reduced set of the solution space.
\item \textbf{The huge solution space:}
The solution space grows exponentially by the number of candidate pragmas. Thus, it is important to alleviate the cost of processing different design configurations.
\item \textbf{Non-smooth impact of design parameters and their correlation:}
For example, changing the unroll factor may not lead to the monotone decrease of the latency~\cite{sohrabizadeh2020autodse, nigam2020predictable}.
\end{itemize}

Separately, current HLS tools optimize the design based on specific code patterns. Although different applications have different domains, they may share the same code structures for some parts. Thus, it is important to identify the different code patterns and learn their effect to be able to transfer the knowledge we gained from one application to another.

In this paper, we aim to address the challenges mentioned above using graph neural network (GNN) with the support for \textit{transfer learning}. For this matter, we developed a framework called {\framework}\footnote{We will open-source the codes once the paper is accepted.} to automate the design process. We first build a model to estimate the different objectives of a design quickly, in milliseconds, without the invocation of the HLS tool. Since the HLS tools employ many heuristics to optimize a design and the design parameters affect each other, we let a deep learning model learn their impact. We represent the program as a graph which includes the program semantics in the form of control, data, call, and pragma flows and exploit a GNN to extract the required features of the graph for predicting the objectives. We propose several techniques for improving the accuracy of the model including Jumping Knowledge Network (JKN)~\cite{xu2018representation}, node attention~\cite{li2015gated}, and multi-head objective prediction. To demonstrate the effectiveness of our model, we build a DSE on top of it for searching through different combinations of the pragmas to find the Pareto-optimal design points. We show that not only can {\framework} find the Pareto-optimal designs for the kernels that were included in its training set, it can also generalize to the kernels outside of its database and detect their Pareto-optimal design points. 
In this paper, we target Xilinx FPGAs as an example but our approach is tool-independent and extendable to Intel FPGAs as well.

In summary, this paper makes the following contributions:
\begin{itemize}
    \item We propose a graph-based program representation for optimizing FPGA designs which includes both the program context and the pragma flow.
    \item We develop a learning model based on graph neural network (GNN) as a surrogate of the HLS tool for assessing a design point's quality in milliseconds order and propose several techniques for improving its accuracy.
    \item We create an automated framework, {\framework}, to build a database of FPGA designs, train a learning model for predicting the design's objectives, and run a design space exploration based on the model to close-in on a high-performance design point.
    \item The experimental results demonstrate that not only can {\framework} find the Pareto-optimal design points for the kernels in its database, it can also optimize the \textit{unseen} kernels by extracting the knowledge it learned from the previously-seen kernels.
\end{itemize}
\section{Background} \label{sec:bg}
\subsection{Programs as Graphs} \label{sec:bg_graph}
A popular way of representing a program as a graph is to extract the \textit{control and data flow graph} (CDFG) of the program from its intermediate representation (IR) in LLVM~\cite{lattner2004llvm}. This way, instead of focusing on the grammar of the code, the semantics of the program flow is captured. In a CDFG, the nodes represent the LLVM instructions that are connected to each other based on the control flow of the program. To represent the data flow of the program, a second type of edge is added between the nodes based on the operands of the instructions. Note that a CDFG includes many low-level operations like memory managment which makes it a desirable representation for FPGA kernels.

\subsection{Graph Neural Networks} \label{sec:gnn}

Recent years have witnessed an increasing amount of graph data which have motivated the researchers to design powerful models for processing graphs. Among various graph machine learning methods, Graph Neural Networks (GNN)~\cite{wu2020comprehensive} are growing in popularity as seen in wide range of applications, e.g. social network analysis~\cite{tan2019deep}, biomedical tasks~\cite{yue2020graph}, traffic forecasting~\cite{jiang2021graph}, etc. 

The core idea of a GNN model is to extract the graph information by learning the features (embeddings) of each node in the graph via aggregating information from its neighboring nodes, commonly referred to as ``message passing''. A GNN model, like a CNN, pass the node embeddings through a series of layers until it derives the rich information for the graph.
The computation of one layer of a typical GNN, in its general form, can be formulated as follows:
\begin{equation}
\label{eq:gnn_general}
 \bm{h}^{\prime}_i = \sigma(\mathrm{TF}(\mathrm{AGG}(\{ \bm{h}_{j} | j \in \mathcal{N}(i) \})))
\end{equation}
where $\bm{h}_{i} \in \mathbb{R}^F$ and $\bm{h}^{\prime}_{i} \in \mathbb{R}^{F^\prime}$ denote the input and output embeddings of node $i$ which are a vector of $F$ ($F^{\prime}$) features, $\mathrm{AGG}$ shows the \textit{aggregation} function which gathers the embeddings of neighbors of each node ($\mathcal{N}(i)$),  $\mathrm{TF}$ represents a \textit{transformation function} to apply on the aggregated results of each node, and $\sigma$ is an activation function to introduce non-linearity to the model.

\begin{figure}[!htb]
	\centering
	\includegraphics[width=0.9\columnwidth]{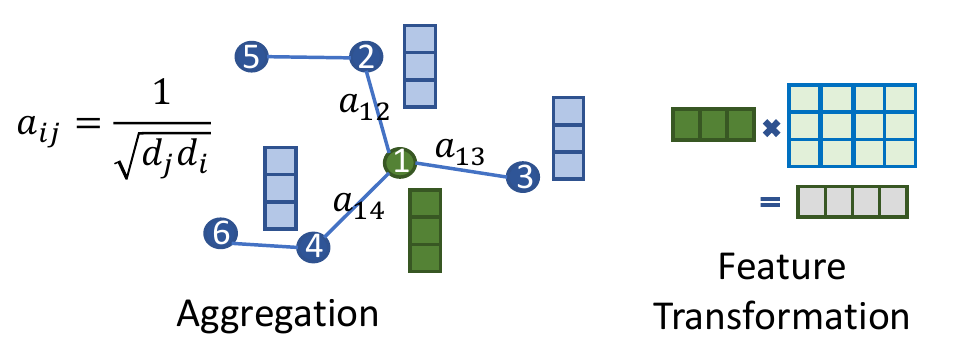} 
	\caption{The computation of a GCN layer}
	\label{fig:gcn}
\end{figure}

\textit{Graph Convolutional Network} (\textsc{GCN})~\cite{kipf2016semi} is a popular form of a GNN which adopts a simple aggregation function that performs a weighted summation of the embeddings of neighboring nodes using the degree of a node, $d_i$, as shown in Fig.~\ref{fig:gcn}:
\begin{equation}
\label{eq:gcn}
\bm{h}^{\prime}_i = \sigma(\bm{W} \sum_{j \in \mathcal{N}(v) \cup
        \{ i \}} \frac{1}{\sqrt{{d}_j {d}_i}} \bm{h}_j)
\end{equation}
where $\bm{W}$ is a trainable weight matrix for the $\mathrm{TF}$ step to act as a filter.
Fig.~\ref{fig:gcn} illustrates this operation for a node. i.e., Node 1. It applies a weighted summation on the embeddings of its neighbors (and itself) based on the $a_{ij}$ values and then multiplies the results with $\bm{W}$ for calculating the output embedding. 

\begin{figure}[!htb]
	\centering
	\includegraphics[width=\columnwidth]{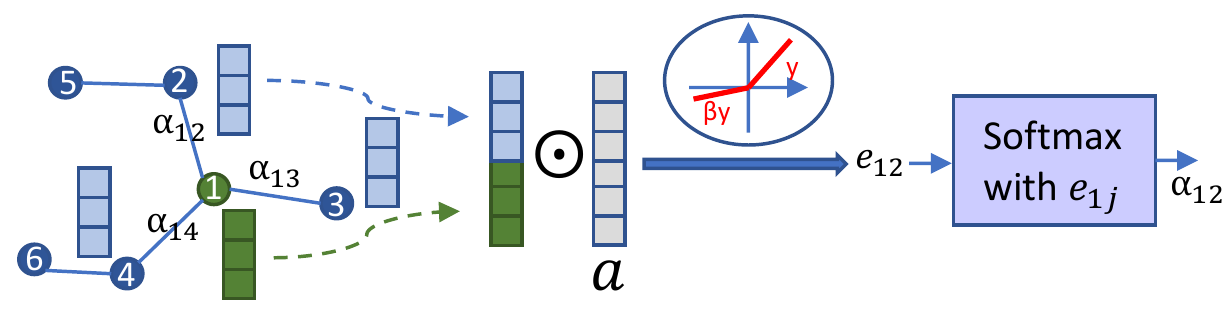} 
	\caption{How the attention coefficient is calculated in a GAT layer.}
	\label{fig:gat}
\end{figure}

One problem with the GCNs is that the aggregation of features of the nodes are based on a fixed set of weights which are determined by the degree of the nodes. Therefore, the model has no way of prioritizing any of the neighbors to learn better embeddings. To solve this problem, another class of GNN models, \textit{Graph Attention Networks} (\textsc{GAT})~\cite{velivckovic2017graph}, were introduced to learn the \textit{importance} of the different neighbors of a node so that they can contribute in updating the node embeddings based on their \textit{attention}. The computation of a GAT layer can be summarized as below:
\begin{equation}
\label{eq:gat}
\bm{h}^{\prime}_i = \sigma(\sum_{j \in \mathcal{N}(i)\cup
        \{ i \}} \alpha_{i,j}\mathbf{W}\bm{h}_{j})
\end{equation}
$\alpha_{i,j}$s are the attention coefficients computed by multi-head dot-product attention. The computation for each head is as follows:
\begin{align}
\begin{split}
\label{eq:gat_att}
         &e_{i,j} = \bm{a}^{\top}
        [\bm{W}\bm{h}_i \, \Vert \, \bm{W}\bm{h}_j] \\
        &\mathrm{LeakyReLU}(y) = max(\beta y, y), 0 < \beta < 1 \\
        \alpha_{i,j} &=
        \frac{
        \exp\left(\mathrm{LeakyReLU}\left(e_{i,j} \right)\right)}
        {\sum_{k \in \mathcal{N}(i) \cup \{ i \}}
        \exp\left(\mathrm{LeakyReLU}\left(e_{i,k}
        \right)\right)} 
\end{split}
\end{align}
where $\Vert$ denotes the concatenation operation and $\bm{a}$ is a learnable vector controlling the attention that node $i$ receives from node $j$. Note that the TF step is the same as GCN and only the AGG step is changed. Fig.~\ref{fig:gat} shows this computation on a toy graph.

\subsection{The Merlin Compiler} \label{sec:merlin}
The Merlin Compiler\footnote{The Merlin Compiler was developed by the Falcon Computing and acquired by Xilinx in late 2020. It is open-sourced at \url{https://github.com/Xilinx/merlin-compiler}}~\cite{merlin, merlin_islped} was developed to make FPGA programming easier by raising its abstraction level.
Inspired by the programming model of OpenMP~\cite{dagum1998openmp}, it introduces a reduced set of high-level compiler directives in the form of pragmas to optimize the design. Based on these pragmas, it performs source-to-source code transformation and automatically generates the respective HLS code along with the required HLS pragmas to enable the designated optimizations. 
Table~\ref{tbl:merlin_pragmas} lists the Merlin Compiler's optimization pragmas. When fine-grained pipelining is enabled (mode \texttt{fg}), the Merlin Compiler tries to pipeline a loop nest while fully unrolling all its sub-loops. The coarse-grained pipelining (mode \texttt{cg}) refers to the case where the Merlin Compiler transforms the code to enable double buffering. Based on these pragmas, the Merlin Compiler automatically employs code transformations to implement memory coalescing, apply memory burst, and cache the required data for enabling the architectural optimizations. 

\begin{table}[!htb]
\centering
\caption{Merlin Pragmas with Architecture Structures}
\label{tbl:merlin_pragmas}
\begin{tabular}{ccc}
\hline
\multicolumn{1}{|c|}{Keyword}                   & \multicolumn{1}{c|}{Available Options}             & \multicolumn{1}{c|}{Architecture Structure} \\ \hline\hline
\multicolumn{1}{|c|}{pipeline}                  & \multicolumn{1}{c|}{mode=cg/fg}                    & \multicolumn{1}{c|}{CG or FG pipelining}   \\ \hline
\multicolumn{1}{|c|}{parallel}                  & \multicolumn{1}{c|}{factor=\textless{}int\textgreater{}}                    & \multicolumn{1}{c|}{CG \& FG parallelism}   \\ \hline
\multicolumn{1}{|c|}{tile}                  & \multicolumn{1}{c|}{factor=\textless{}int\textgreater{}}                    & \multicolumn{1}{c|}{Loop Tiling}   \\ \hline
\multicolumn{3}{c}{CG: Coarse-grained; FG: Fine-grained}
\end{tabular}
\end{table}

We chose to utilize the Merlin Compiler as the backend of our tool to not only reduce the size of the solution space, but also exploit its code transformations to achieve a better design. Nonetheless, by employing the Merlin Compiler, the GNN model works harder to learn because it needs to decipher when (and where) the Merlin Compiler applies its automated optimizations.

\section{Problem Formulation} \label{sec:problem}
In this work, we aim to speed up the DSE problem for HLS. For this matter, we first seek to develop a prediction model that can mimic the HLS tool by estimating the quality of design. Then, using our predictive model, we run a DSE on different design parameters (pragmas) to find out the best combination of parameters for optimizing the design. More formally, we propose solutions for the following two problems:

\noindent\textit{\textbf{Problem 1: Build the Prediction Model.}} Let $\mathcal{P}$ be a C program as the FPGA accelerator kernel with $K$ parameters ($p_i$) denoting different possible HLS pragma candidates as design configurations ($\theta$), where $\mathbb{R}^{K}_{\mathcal{P}}$ is a set of all the different configurations:
\begin{align}
    \theta = [p_0, p_1, ..., p_K] \in \mathbb{R}^{K}_{\mathcal{P}}
\end{align}
Let $\mathbf{H}$ be a vendor HLS tool that outputs the true execution cycle $Cycle(\mathbf{H},\mathcal{{P}(\theta)})$ and the true resource utilization $Util(\mathbf{H},\mathcal{{P}(\theta)})$ of the program $\mathcal{P}$:
\begin{align}
    \mathbf{Q_H}(\mathbb{R}^{K}_{\mathcal{P}}) = \big\{ \big( Cycle(\mathbf{H},\mathcal{P}(\theta)), Util(\mathbf{H},\mathcal{P}(\theta)) \big) \big| \theta \in \mathbb{R}^{K}_{\mathcal{P}}\big\}
\end{align}
Find a prediction function ($\mathbf{F}$) that approximates the results of $\mathbf{H}$ for any given program $\mathcal{P}$ with any design configurations ($\theta$) having any number of parameters:
\begin{align}
\underset{\mathbf{F}}{\min}\ \, Loss(\mathbf{Q_F}(\mathbb{R}^{K}_{\mathcal{P}}),\, \mathbf{Q_H}(\mathbb{R}^{K}_{\mathcal{P}}))
\end{align}
In case of a regression task, the loss function is calculated using root mean squared error (RMSE) over all the designs. For the classification task, the percentage of misclassified cases are considered.

\noindent\textit{\textbf{Problem 2: Identify the Optimal Configuration.}} Given a C program $\mathcal{P}$ as the FPGA accelerator kernel with $K$ parameters along with its design space set $\mathbb{R}^{K}_{\mathcal{P}}$, and a prediction function $\mathbf{F}$, find a configuration $\theta \in \mathbb{R}^{K}_{\mathcal{P}}$ in a given search time limit so that the generated design $\mathcal{P}(\theta)$ can fit in the FPGA and the execution cycle is minimized. Formally, our objective is:
\begin{align}
\underset{\theta}{\min}\ Cycle(\mathbf{F}, \mathcal{P}(\theta))
\end{align}
\vspace{-0.3cm}
\noindent subject to
\begin{align}
\begin{split}
\theta \in \mathbb{R}^{K}_{\mathcal{P}}, \quad
\forall u \in Util(\mathbf{F}, \mathcal{P}(\theta)), u < T_u\
\end{split}
\end{align}

\noindent where $u$ is the utilization of one type of the FPGA on-chip resources and $T_u$ is a user-defined threshold for that type on the FPGA. 

\section{Related Work} \label{sec:rel_work}
With FPGA's synthesis time being a huge bottleneck in its development cycle, a new research domain has been made to explore the solution space more efficiently. Because of the unpredictability of the HLS tool, they treat it as a black-box and invoke the tool each time they want to get a feedback on the quality of design~\cite{s2fa, schafer2017parallel, sohrabizadeh2020autodse}. To explore the space they either make use of the general search heuristics (such as greedy mutation, simulated annealing, and genetic algorithm), or develop their own heuristic which is more suitable for the HLS DSE problem. However, their performance is limited by the runtime of the HLS tool; hence, they can explore only a small fraction of the design choices.

To speed up the search, a number of previous works have developed a model to employ instead of the invocation of the HLS tool. 
The authors in~\cite{comba, linanalyzer} propose an analytical model to predict the design objectives based on the program's dependence graph. Their models are independent of the HLS tool and rely on only the behavioral description of the program. This can potentially harm the model's accuracy as the HLS tools utilize multiple heuristics that can lead to different results than the theoretical ones, so they should also be modeled~\cite{dse-survey}. To improve the accuracy of the model, a category of the previous studies restrict their application domain to limit their accelerators to either a particular application, a well-defined microarchitecture template, or a specific computation pattern~\cite{autoaccel, zacharopoulos2019compiler, autosa, chi2018soda, sohrabizadeh2020end, xu2020autodnnchip, zheng2020flextensor}. The downside is that they lose generality and are not applicable to other domains.

\begin{figure*}[!htb]
	\centering
	\includegraphics[width=\linewidth]{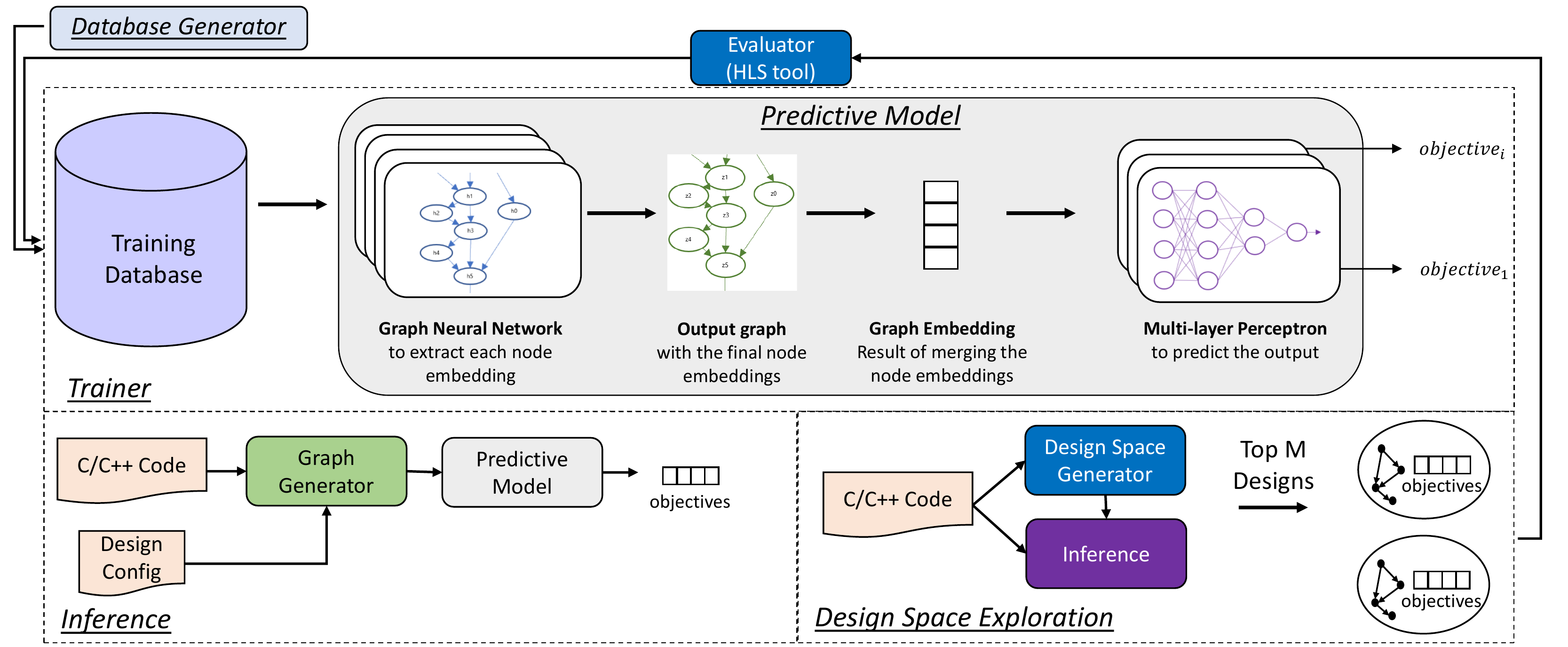} 
	\caption{High-level overview of the {\framework} framework}
	\label{fig:framework}
\end{figure*}

A series of other works use a learning algorithm to resemble the HLS tool and predict the design's objectives. These works iteratively sample the solution space, synthesize the designs using the HLS tool, and adopt a supervised learning algorithm (such as random forest or decision tree) until the model gets to an acceptable accuracy~\cite{liu2013learning, mahapatra2014machine, meng2016adaptive}. The drawback of these approaches is that they build a separate learning model per application and the results from one application are not transferred to the other one. Kwon et al.~\cite{kwon2020transfer} propose a transfer learning approach using a multi-layer perceptron (MLP) network to solve this problem. As the input to the model, they only use the pragma configurations. However, as we shall show in Section~\ref{sec:eval_model}, not taking the program semantics into account harms the accuracy significantly.

A few of the very recent works have proposed to use GNN for predicting the design quality~\cite{wu2021ironman, ustun2020accurate}. Ustun et al.~\cite{ustun2020accurate} proposes a GNN-based model to learn the operation mapping to FPGA's resources for delay prediction in HLS. IronMan~\cite{wu2021ironman} uses GNN to predict the performance of the program under different resource allocations (DSP or LUT) to the computation nodes. Yet, none of these works include the pragmas in their input representation so their models cannot be used for finding the best design configuration. This paper is the first work to employ a graph representation that captures both the program semantics and the pragmas. Furthermore, we show that our approach is capable of learning from a set of applications, building a \textit{single} predictive model for all of them, and extending its knowledge to new applications.

\section{Our Proposed Methodology} \label{sec:approach}
Fig.~\ref{fig:framework} depicts a high-level overview of {\framework} which operates in three modes: training, inference, and DSE. When training is enabled, {\framework} takes its \textit{training database} as the input and trains a \textit{predictive model} to learn to estimate the design's objectives. The training database contains designs with various configurations from different applications along with their synthesized results using the HLS tool as the true value of the their objectives (Section~\ref{sec:db}). 

Instead of manually analyzing the designs and extracting the features that are useful for estimating the quality of design, we let a neural network model discover them since it is less prone to error and can potentially capture the flaws of the HLS tool. To do this, {\framework} represents each design in the database as a graph (Section~\ref{sec:graph}). Then, the {\framework}'s predictive model assembles a GNN model to learn a graph embedding to be fed into a multi-layer perceptron (MLP) network for estimating the different objectives of the design. When the GNN and MLP are trained together, the model tries to adjust the graph embedding in a way that it includes the highest information for predicting the objectives (Section~\ref{sec:model}).

Once the \textit{trainer} has optimized the predictive model, it can be used for the \textit{inference} stage, where {\framework} gets a C/C++ code as an input along with the desired design configuration. It then adopts the {\framework}'s \textit{graph generator} to get the graph representation of the design. Finally, it employs the trained predictive model to estimate the design's objectives. During the DSE phase, each of the design configurations are evaluated using the predictive model as in the inference stage. {\framework} continuously accumulates more design points; thus, once the DSE is finished, the top M designs are synthesized using the HLS tool to augment the database with their \textit{true} objectives. This can help us include better representatives of the space in our database (Section~\ref{sec:dse}).

\subsection{Database Generation} \label{sec:db}
\begin{figure}[!htb]
	\centering
	\includegraphics[width=\columnwidth]{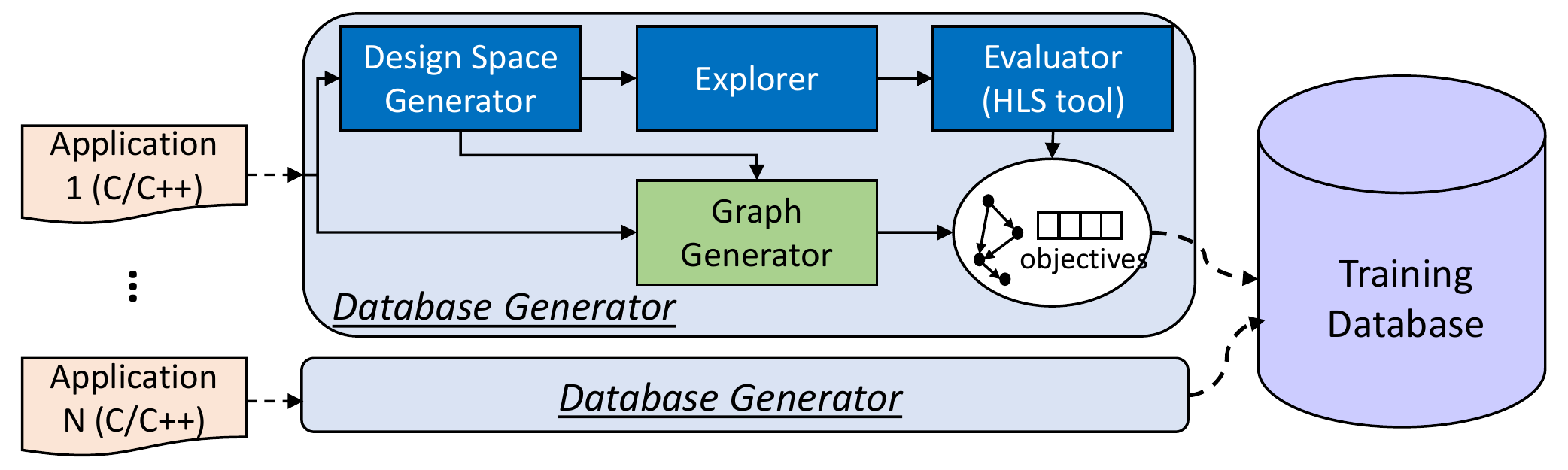} 
	\caption{Database generator of {\framework}}
	\label{fig:db-gen}
\end{figure}

We adapt a related prior work, AutoDSE~\cite{sohrabizadeh2020autodse}, to generate the initial database for each of the applications. AutoDSE is also built on top of the Merlin Compiler and, by default, it adopts a bottleneck-based optimizer to reduce the number of iterations needed to close-in on high-performance design points. Fig.~\ref{fig:db-gen} demonstrates our approach for generating the database. We follow the same rules as AutoDSE in the \textit{design space generator} for defining the solution space and augmenting the input code with candidate pragmas. More specifically, each \texttt{for} loop can take three pragmas: \texttt{pipeline}, \texttt{parallel}, and \texttt{tile} except for where a pragma is not applicable (e.g., the \texttt{tile} pragma can only be applied when the loop has an inner loop). Note that the Merlin Compiler automatically inserts the rest of the HLS pragmas (such as \texttt{array\_partition}) that are needed to satisfy these optimizations. We also make use of the rules AutoDSE has for pruning a design configuration (e.g., when fine-grained pipelining is applied on a loop, all the pragmas for the inner loops should be set to the default value).

The optimizer in AutoDSE is intended to only explore those designs that are promising to have a high quality, i.e., ``good'' design configurations. However, this is not enough if we want to train a model to predict the results of the HLS tool. In other words, we need to have a variety of design points from ``bad'' to ``good'' so that the model can learn to distinguish them. Therefore, {\framework}'s \textit{explorer} extends AutoDSE to exploit three types of explorers:
\begin{itemize}
    \item The existing explorer of AutoDSE, bottleneck-based optimizer, which can find high quality designs.
    \item A hybrid explorer that is a combination of the bottleneck-based optimizer and exhaustive search. It first uses the bottleneck-based optimizer to improve the design quality. When the quality of the best design point is improved by $X\%$, it will evaluate up to $P$ neighbors of the new design point (both $X$ and $P$ are user-defined variables). A neighbor is a point that the option for only one of its pragmas is different from the current point. By employing this explorer, the model can see how a design's quality is changed by modifying only one of its pragmas.
    \item A random explorer which may consider those configurations that the previous two explorers would skip.
\end{itemize}

Once the explorer selects a design point for evaluation, it is passed to the Merlin Compiler which generates the equivalent HLS code with the HLS pragmas and invokes the HLS tool to get the design's objectives. In parallel, the \textit{graph generator} (Section~\ref{sec:graph}) builds the graph representation of the program with that design configuration. Finally, the graph representation of the program, along with the HLS results, are committed to a common database. The \textit{training database} gradually collects results from \textit{different} applications in a shared space to be used for training the model.

\subsection{Program Representation} \label{sec:graph}
The first step in building a predictive model is to define a representation of the program which not only captures its behavioral flow, but also includes the pragmas. 
As mentioned in Section~\ref{sec:bg_graph}, CDFG is a popular choice for this purpose. 
On the downside, when building the graph, the CDFGs ignore the precision of the operands and their values, which are crucial in determining design's objectives. Recently, a newer program representation is proposed, PrograML~\cite{programl}, which extends the CDFG by explicitly assigning separate nodes to operands to retrieve the missing information. 
PrograML is a language-independent representation of the program semantics that takes the LLVM IR~\cite{lattner2004llvm} of the design as the input and builds a directed graph containing its control, data, and call flow. As such, we build our program representation by extending PrograML and including the \textit{pragma} flow. 

In {\framework}, each of the candidate pragmas are defined in either of the following forms:
\vspace{0.05in}
\begin{lstlisting}[numbers=none]
#pragma ACCEL pipeline auto{pragma_name}
#pragma ACCEL parallel factor=auto{pragma_name}
#pragma ACCEL tile factor=auto{pragma_name}
\end{lstlisting}

\noindent where the \texttt{pragma\_name} is a placeholder for the variable storing the pragma's option. The choices for \texttt{pipeline} pragma are \texttt{(off|cg|fg)} where \texttt{off} disables the pragma. The other two pragmas take a numerical value as their option as mentioned in Table~\ref{tbl:merlin_pragmas}. When their factor is set to $1$, they will be disabled. 

\begin{figure}[!htb]
	\centering
	\includegraphics[width=0.7\columnwidth]{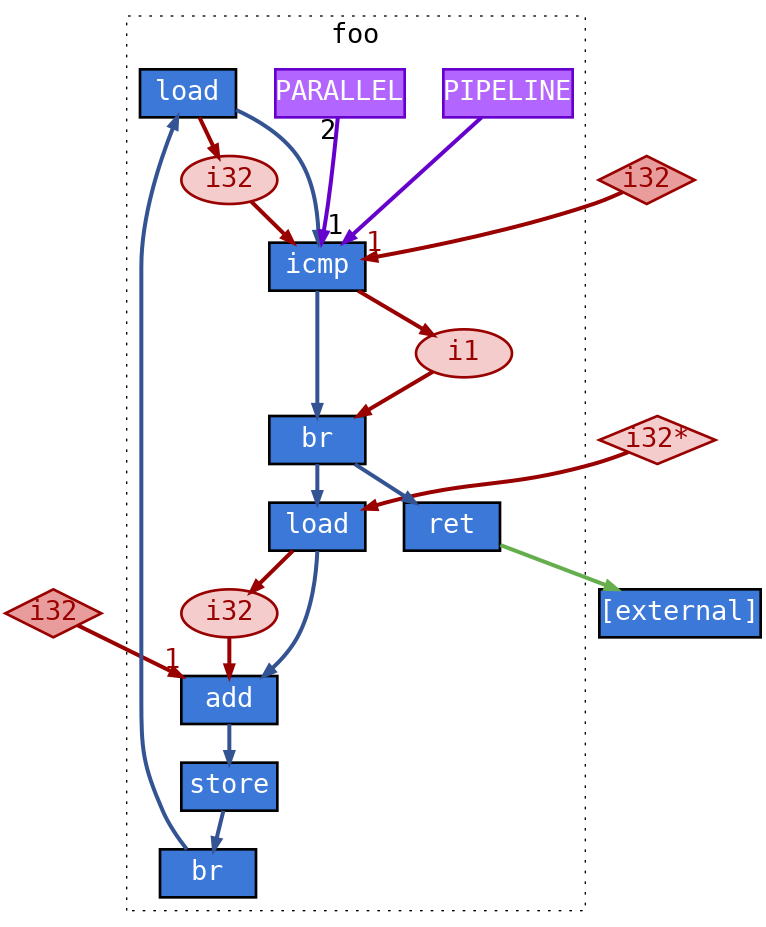} 
	\caption{The graph representation of Code~\ref{code:simple-for} illustrating the different kinds of nodes and edges in our representation.}
	\label{fig:graph}
\end{figure}

For each candidate pragma, we augment the graph generated by PrograML with a node that stores the placeholder pragma. Since the pragmas are applied to the loops, we connect this node to one of the instruction nodes corresponding to the loop: \texttt{icmp}. Code~\ref{code:simple-for} shows the code snippet for a toy example having a simple \texttt{for} loop with two candidate pragmas. Fig.~\ref{fig:graph} depicts its graph representation.

\begin{lstlisting}[caption=Code snippet of an input toy example to {\framework} \label{code:simple-for},
   floatplacement=H]
void foo(int input[N]) {
#pragma ACCEL pipeline auto{_PIPE_L1}
#pragma ACCEL parallel factor=auto{_PARA_L1}
  for (int i = 0; i < N; i++) {
        input[i] += 1; } } 
\end{lstlisting}

As Fig.~\ref{fig:graph} demonstrates, there are three types of nodes in each graph. The first kind (the blue ones) contains the LLVM instructions that together demonstrate the control flow of the program. The second kind (the red ones) exhibits the constant values and variables that capture the data flow of the program.
Note that we have removed some of the instruction and data nodes and kept only the relevant nodes to shrink the graph for illustration purposes.
The pragma nodes are presented as purple boxes connecting to the respective \texttt{icmp} node. In this graph, in addition to the nodes, the edges also have different kinds which show the different flows of the graph: control (blue), data (red), call (green), and pragma (purple). When there are two or more edges of the same type connected to a node, they are numbered to further distinguish them (see the edges connecting from pragma nodes to the \texttt{icmp} node).

For nested loops, to accurately distinguish the different candidate pragmas, one must know the level of loop to which the pragmas correspond. As a rule of thumb, the HLS tools perform better when the pragmas are applied to inner loops since they can implement fine-grained optimizations easier~\cite{sohrabizadeh2020autodse}. Fortunately, our graph representation contains the loop level since it includes the control flow of the program. It can also be encoded in each node using the LLVM block ID of the \texttt{for} loop. More specifically, each of the nodes and edges have the following attributes:
\vspace{0.05in}
\begin{lstlisting}[numbers=none]
Node = {'block': LLVM block ID, 'full_text': Full description of the node, 'key_text': Node key task, 'function': Function ID, 'type': Node type}
Edge = (Src node ID, Dst node ID, {'flow': Flow type, 'position': Position ID})
\end{lstlisting}
\noindent where the \texttt{type}, \texttt{flow}, and \texttt{position} attributes encode this information:
\begin{center}
\footnotesize
\begin{tabular}{c||c|c|c|c}
\textbf{type} & 0: instruction & 1: variable & 2: constant value & 3: pragma\\ \hline
\textbf{flow} & 0: control & 1: data & 2: call & 3: pragma\\ \hline
\textbf{position} & 0: tile & 1: pipeline & 2: parallel & -\\ 
\end{tabular}
\end{center}

\begin{figure}[!htb]
	\centering
	\includegraphics[width=\columnwidth]{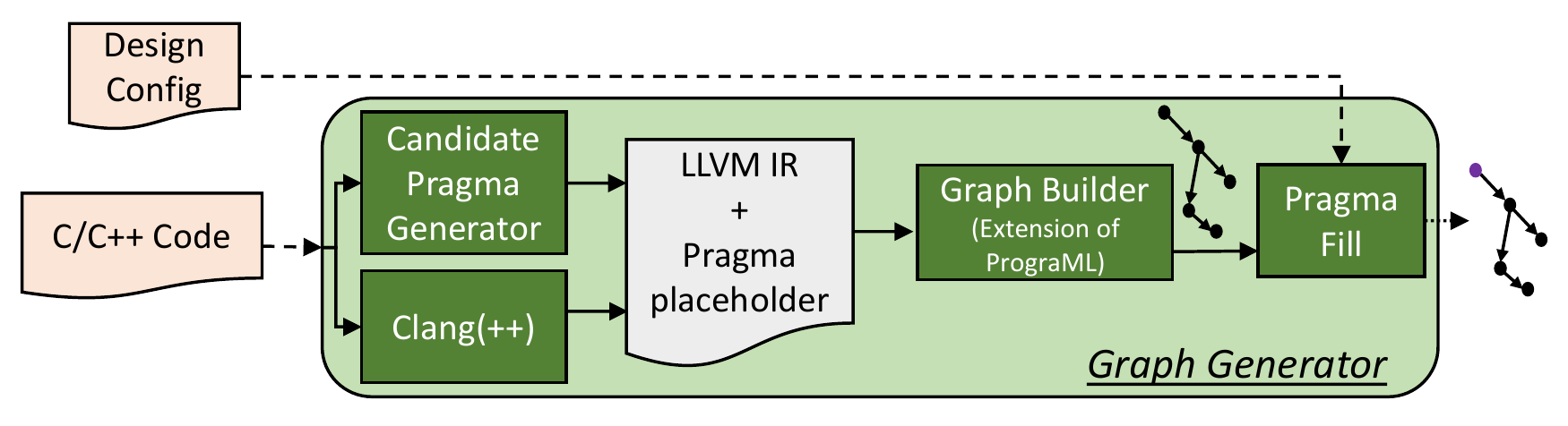} 
	\caption{Graph generator of {\framework}}
	\label{fig:graph-gen}
\end{figure}

The \texttt{full\_text} attribute stores the complete form of the text each node represents, while the \texttt{key\_text} attribute shows a keyword corresponding to that node. Here is an example for each of the control, data, and pragma nodes from the graph in Fig.~\ref{fig:graph}:
\vspace{0.05in}
\begin{lstlisting}[numbers=none]
'full_text': #pragma ACCEL PIPELINE auto{_PIPE_L1}, 'key_text': PIPELINE
'full_text': %0 = load i32, i32* %i, align 4, 'key_text': load
'full_text': i32* %input, 'key_text': i32*
\end{lstlisting}

Fig.~\ref{fig:graph-gen} depicts the flow of the \textit{graph generator} in {\framework}. Its input is a C/C++ code, and it first generates the LLVM IR representation of the program, along with its candidate pragmas which are produced based on the same rules as the \textit{design space generator} in Section~\ref{sec:db}. Then, the \textit{graph builder} outputs their respective graph. For each design configuration, the \texttt{auto} variables in the pragma placeholders are replaced with their corresponding values. Therefore, among the graphs for different design configurations of the same application, only the attributes of their pragma nodes are different.

\subsection{Predictive Model} \label{sec:model}
\begin{figure*}
	\centering
	\includegraphics[width=0.9\linewidth]{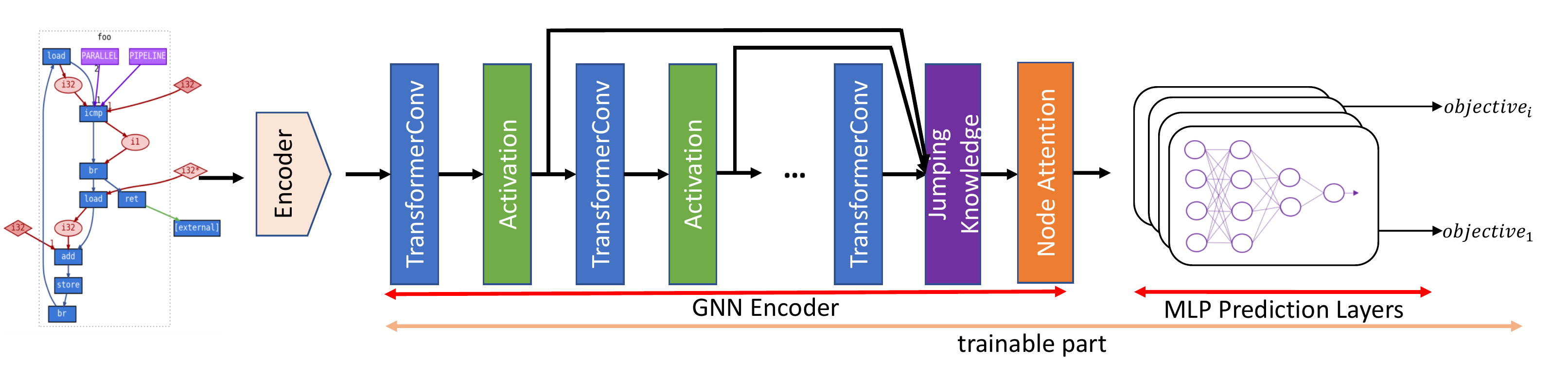} 
	\caption{The architecture of {\framework} predictive model.}
	\label{fig:model}
\end{figure*}

Fig.~\ref{fig:model} depicts our model architecture for predicting the design's objectives. As the figure shows, it starts with taking the graph representation of the program as the input. Then, it encodes the nodes' and edges' attributes (Section~\ref{sec:graph}) to create their initial embeddings. To do this, we go over all the design points in our database, build a list of all the available attributes along with the available options for each of the pragmas, and create a one-hot encoding for each of them. Then, for each node/edge, we go over their attributes, encode them, and concatenate them to create the initial embeddings. This encoding helps the model to assign a higher weight to the attributes that contribute more to the prediction of final objectives. 
Once the initial embeddings are available, the model exploits a \textit{GNN encoder} (Section~\ref{sec:gnn-encoder}) to learn which information of the graph to extract for predicting the design's objectives. After which, the GNN encoder passes the graph embeddings to a set of MLPs to estimate the outputs (Section~\ref{sec:MLP}).

\subsubsection{GNN Encoder:} \label{sec:gnn-encoder}
The GNN encoder transforms a graph $\mathcal{G}$ with its initial embeddings into a $D$-dimension embedding, $\bm{h}_{\mathcal{G}} \in \mathbb{R}^D$. This encoder consists of three stages: (1) sequentially stacked \tconv layers which produce node embeddings, (2) a Jumping Knowledge Network which combines the node embeddings from different layers to produce the final node embeddings with dynamic ranges of neighborhoods, and (3) an attention mechanism to aggregate the node-level embeddings to produce the final graph-level embedding.

\textbf{\tconv} \enspace
We reviewed two well-known types of GNN layers for building the node embeddings in Section~\ref{sec:gnn}: GCN~\cite{kipf2016semi} and GAT~\cite{velivckovic2017graph}. One drawback of these layers is that they both overlook the edge embeddings. \tconv~\cite{transconv_shi2020masked} is a state-of-the-art GNN architecture, which combines Transformer~\cite{vaswani2017attention,devlin2018bert} with message passing to update a node's embedding. Similar to GAT, it builds attention coefficients ($\alpha_{i,j}$) for aggregating the neighbors, but in a different manner inspired by the Transformer model and including the edge embeddings:
\begin{equation}
\label{eq:tconv_att}
\alpha_{i,j} = \textrm{softmax} \left(
    \frac{(\mathbf{W}_1\bm{h}_i)^{\top} (\mathbf{W}_2\bm{h}_j + \mathbf{W}_3\bm{e}_{ij})}
    {\sqrt{D}} \right)
\end{equation}
where $\bm{W}_1$, $\bm{W}_2$, and $\bm{W}_3$ are learnable weight matrices, and $\bm{e}_{ij}$ denotes the embedding of the edge between nodes $i$ and $j$. The fact that it includes the edge attributes in the formulation of the $\alpha_{i,j}$s is a desirable feature for our task since the edges in our graph representation contain useful information (Section~\ref{sec:graph}) we would like to utilize when updating the node embeddings. In addition, \tconv makes use of gated residual connections when updating the node embeddings that can prevent the model from over-smoothing. Consequently, we adopt \tconv as the basic building block of our model. The evaluation results in Section~\ref{sec:eval_model} show that the \tconv indeed reduces the loss by $1.48\times$ under the same number of learning iterations.

\textbf{Jumping Knowledge Network} \enspace 
Each layer of a GNN gathers the embeddings of the first-order neighbors. By adding each layer, the nodes will receive the embeddings from one hop further since their first-order neighbors are now updated with their own first-order neighbors. The different nodes in the graph may need different ranges of neighborhoods for updating their embeddings. For example, in the graph depicted in Fig.~\ref{fig:graph}, both the \texttt{load} and \texttt{add} nodes are affected by the pragma nodes. The \texttt{load} node sees the effect of the pragma node after 3 layers, while the \texttt{add} node is impacted after 4 layers. By employing 4 layers of GNN for both of these nodes, the \texttt{load} node may receive some extra noise from new nodes which can affect its embeddings negatively.

Therefore, to fully leverage the embeddings generated by different layers of the GNN model, we exploit the Jumping Knowledge Network (JKN)~\cite{xu2018representation} which as Fig.~\ref{fig:model} illustrates, takes in the output of all the layers to flexibly pick different ranges of neighborhood for each node. A JKN, in its simplest form, exploits max pooling for choosing the final embeddings of each node among its node embeddings from all the previous GNN layers:
\begin{equation}
\label{eq:jkn}
\bm{h}_i =         \max \left( \bm{h}_i^{(1)}, \ldots, \bm{h}_i^{(T)} \right)
\end{equation}
where $\bm{h}_i^{(k)}$ denotes the embedding of node $i$ after the $k$-th layer.

\textbf{Node attention-based graph-level embedding generation:} \enspace To generate one vector representation for the entire graph, one can simply perform a global summation of all the node embeddings:
\begin{equation}
\label{eq:sum_g_emb}
        \bm{h}_{\mathcal{G}} = \sum_{i=1}^{N} \bm{h}_i.
\end{equation}
where $\bm{h}_{\mathcal{G}}$ and $N$ denote the graph-level embedding and the number of nodes in the graph, respectively.
However, given the fact that our graph representation contains both the pragma nodes and the program context nodes, it is preferable to introduce attention that learns which node is more important for the prediction tasks. Therefore, we adopt the node attention mechanism proposed in \cite{li2015gated} for producing the graph-level embedding: 
\begin{equation}
\label{eq:att_g_emb}
        \bm{h}_{\mathcal{G}} = \sum_{i=1}^{N} \mathrm{softmax} \left(
        \textrm{MLP}_1 ( \mathbf{h}_i ) \right) \cdot
        \textrm{MLP}_2 ( \mathbf{h}_i )
\end{equation}
where $\textrm{MLP}_1$ maps the node embedding from $\mathbb{R}^D$ to $\mathbb{R}$ followed by a global softmax to obtain one attention score per node. The attention scores are then applied to the transformed node embeddings, $\textrm{MLP}_2 ( \mathbf{h}_i )$, to obtain the final graph-level embedding.

\begin{figure}
\centering
\includegraphics[width=0.85\columnwidth]{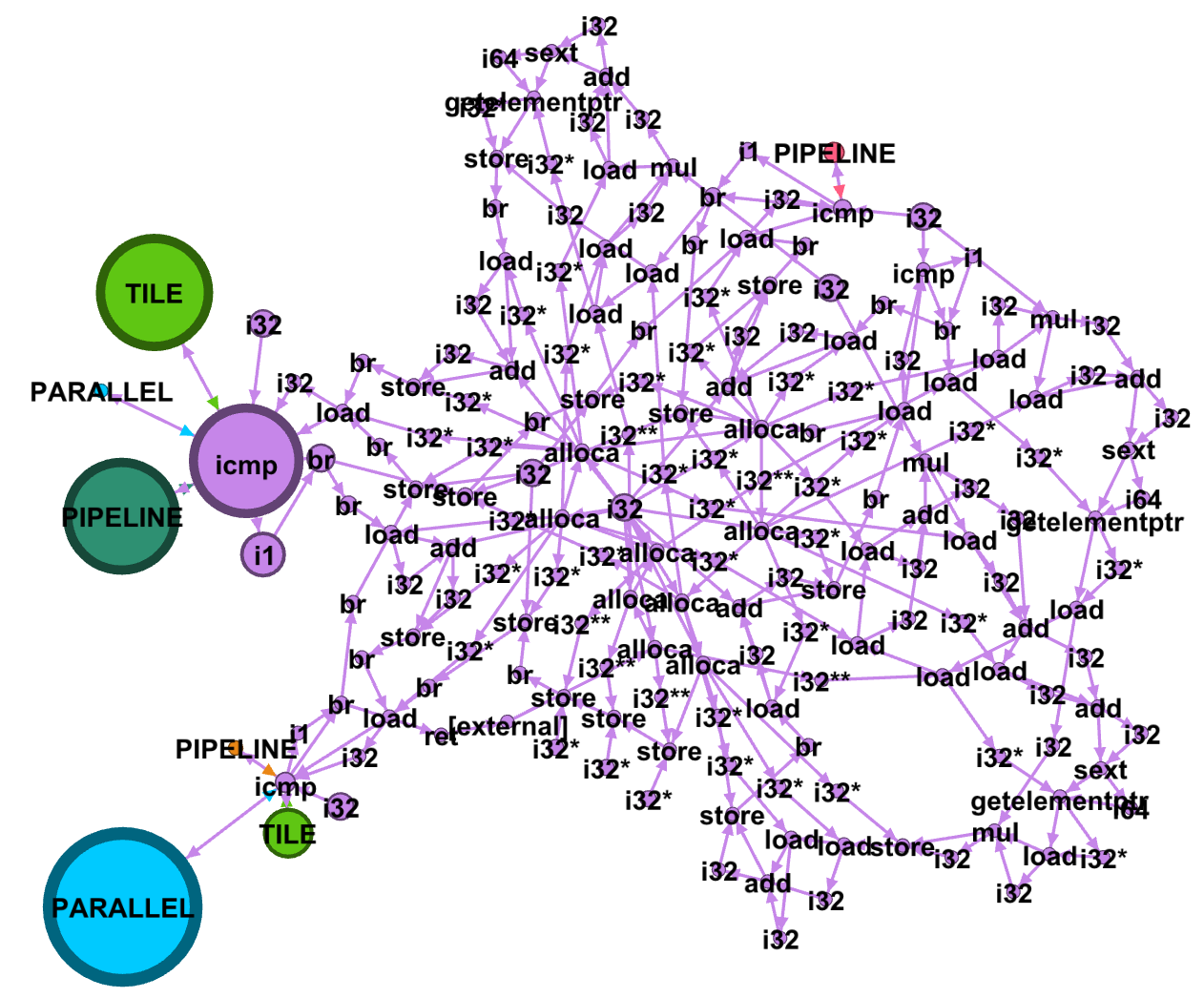}
\caption{Visualization of the node attention scores of a design of the stencil kernel. The larger the circle, the higher its attention is. The plot is created using \textsc{Gephi}~\cite{bastian2009gephi}.
}
\label{fig:stencil-node-att}
\end{figure}

Fig.~\ref{fig:stencil-node-att} depicts the graph for a design of the stencil (stencil2d) kernel in Machsuite benchmark~\cite{machsuite}. The size of the circle for each node is proportional to the attention that its node embedding receives in building the graph-level embedding. As we expected, the pragma nodes are among the most \textit{important} nodes. Yet, the model could learn that not all the pragma nodes are equally important. As the figure suggests, the loop trip count (\texttt{icmp} node and \texttt{i32} node connecting to it) and other contextual information of the loop determines the importance of the loop's pragmas. The results show that, the model could correctly learn that for the outer-most loop level, \texttt{pipeline} and \texttt{tile} pragmas are more important whereas, for the loop inside it, \texttt{parallel} is the most dominant pragma.

\begin{figure*}
     \centering
     \subfloat[Designs represented by the initial features.]{\includegraphics[width=0.45\textwidth]{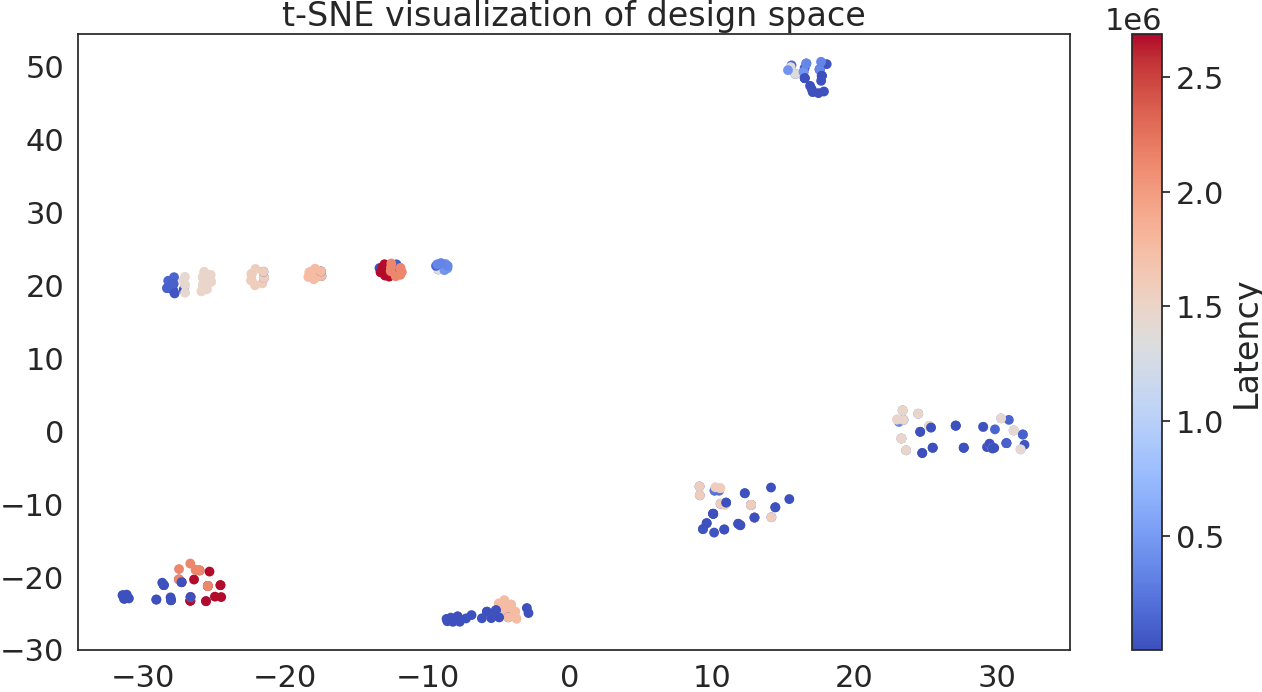}}
     \subfloat[Designs represented by the embeddings learned by \model.]{\includegraphics[width=0.45\textwidth]{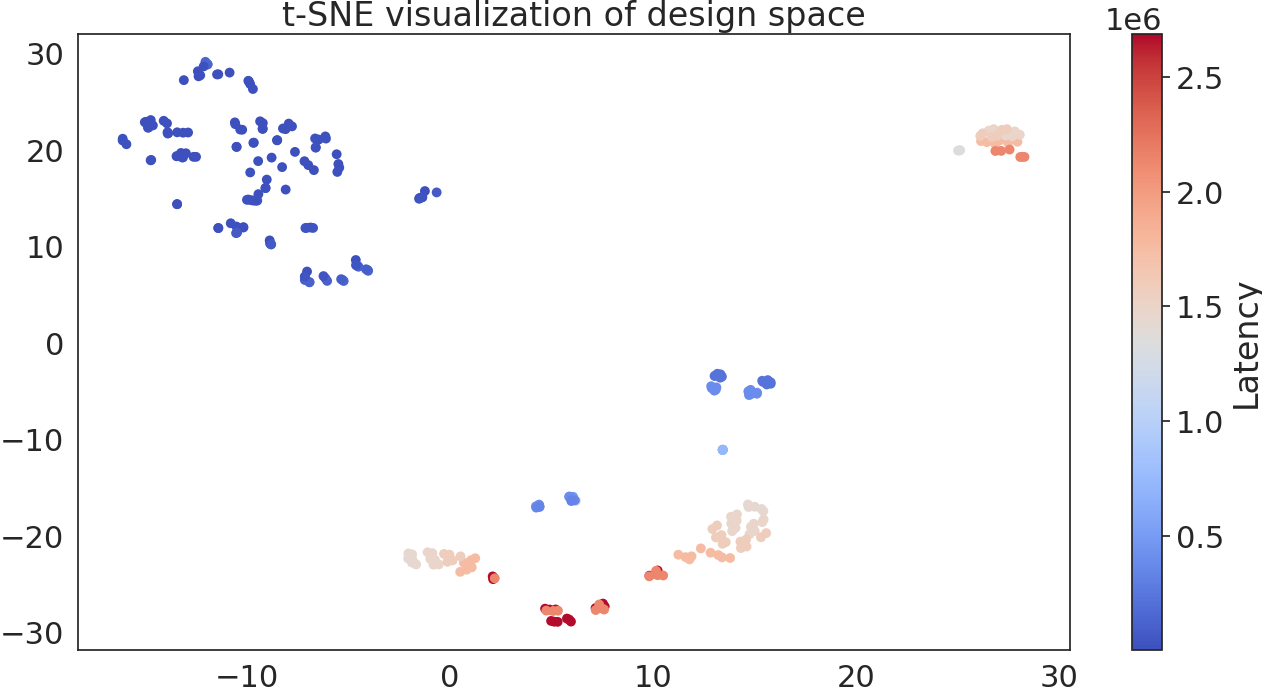}}
     \caption{Visualization of the design configurations of stencil via t-SNE~\cite{maaten2008visualizing}. Each point represents a design with colors indicating its latency value. 
     }
     \label{fig:tsne_stencil}
\end{figure*}

The initial node embeddings in our experiments (which we created by concatenating the encoded version of the node attributes) are 124-D vectors and the final embeddings of our GNN encoder are 64-D vectors. We utilize t-SNE~\cite{maaten2008visualizing} to visualize these vectors by down-projecting them into 2-D space. t-SNE is a powerful technique that can model high-dimensional data by 2-D points in a way that nearby (distant) points model similar (dissimilar) data.
Fig.~\ref{fig:tsne_stencil}(a) depicts the t-SNE plot for the stencil kernel based on its initial embeddings. For each design, we create a graph-level embedding by adding the nodes' initial embeddings. Each point in the plot corresponds to a design which is color-coded by the design's latency (cycle counts). Fig.~\ref{fig:tsne_stencil}(b) demonstrates how the t-SNE plot changes if we use the graph-level embeddings generated by our GNN encoder instead. As the figures suggest, while the initial features show high similarity between two design points with a huge difference in their latency value, the GNN encoder could successfully distinguish them and assign embeddings to the graphs in a way that only the designs with similar latency be clustered together.

\subsubsection{MLP Prediction Layers:} \label{sec:MLP}
After the GNN encoder has encoded the graph into a $D$-dimensional representation, further transformation is needed to perform the final prediction task. We have the following two learning tasks for assessing a design point:
\begin{itemize}
    \item \textbf{Classification task for determining whether a design configuration is valid.} First, we check whether the combination of the pragmas create a valid design configuration or not. The source of invalidity can come from many factors such as: 1) the pragma combinations may create a design that is hard for the HLS tool to optimize. If synthesis of a design does not finish in 4 hours, we mark it as invalid; 2) the HLS tool may refuse to synthesize a design due to the use of combined high parallelization factors; 3) a combination of the pragmas may not be feasible in general. For example, the coarse-grained pipelining is implemented by applying double buffering on the loop, which is not parallelizable since at each iteration a batch of data should be transferred from DRAM to BRAM. When an optimization is not applicable, the Merlin Compiler produces a warning for it which we use to mark the design as invalid. Although AutoDSE~\cite{sohrabizadeh2020autodse} identities some of the invalid cases (which we use), it does not cover all of the cases so we let the model learn them.
    \item \textbf{Regression task to estimate the design's objectives.} Once we identify that a design is valid, we estimate the quality of design by predicting its cycle count and resource utilization which we define as our regression task. 
\end{itemize}
For each of these tasks, we exploit an MLP to do the prediction based on the graph-level embeddings. Note that our regression task is seeking to predict multiple objectives. To achieve this, we can either employ separate models for each of them or use the same GNN encoder as the backbone and adopt a different MLP branch for each target objective, as seen in Figure~\ref{fig:model}. In the former, each model tries to find the graph-level embedding needed to predict the target objective separately. However, in the latter, they can help each other in extracting the useful graph information for predicting the outputs since the same graph-level embedding is fed into multiple branches. This is a desirable feature for us as our objectives are correlated with each other. 

\subsection{Design Space Exploration} \label{sec:dse}
After building an accurate model for assessing a design point, we can search through the different design choices to pick the best one. For each design configuration, we first run the classification model to determine whether it is valid or not. If it is valid, we then run our regression models to predict the design's objectives. If, for any of the resources, the utilization is higher than 0.8 ($80\%$), we reject that design due to over-utilization. $80\%$ is an empirical threshold which, when the utilization exceeds it, the design will suffer from frequency degradation and difficulty in mapping. Finally, among the remaining designs, we pick the top 10 designs (with the least latency numbers) to be evaluated using the HLS tool. This means that our model helped us to run the HLS tool only 10 times instead of invoking it for every design. 

Since our models can finish in milliseconds, we can explore a large number of design points very quickly. Nevertheless, for enormous solution spaces, we still may not be able to search through the whole space in a timely fashion. Therefore, we set a time limit for running the DSE and employ a heuristic to prioritize searching through the most-promising candidates first. As the HLS tools can implement the fine-grained optimizations better, we sort the pragmas by their loop level so that the pragmas of the inner-most loop levels are evaluated sooner. 

For ordering the pragmas, starting with the inner-most loops, we adapt a BFS-like traversal of their pragmas. More specifically, starting from the inner-most loop level of each of the nested loop sections, we pick a pragma to explore first. If there are more than one pragma at that level, we prioritize \texttt{parallel} over \texttt{pipeline} over \texttt{tile}. If the picked pragma has a dependency from the same loop level or one loop level further, we prioritize evaluating that dependency before other pragmas. A pragma has a dependency when we want to prune the invalid design combinations. For example, there is always a dependency between the \texttt{parallel} pragma of one loop level with the \texttt{pipeline} pragma of its upper loop level. This is because \texttt{fg} pipelining completely unrolls the sub-loops so we no longer need the \texttt{parallel} pragma. After evaluating this pragma, we do the same process for the next loop section. We continue to do this until all the pragmas are added to our ordered list. Since there is always a dependency between the \texttt{parallel} pragma of one loop level with the \texttt{pipeline} pragma of its upper level, for the second-inner-most loop level upwards, this ordering results in evaluating the pipeline pragma first before any other optimizations (even before pipelining or tiling of its inner loop). This is desirable since, if this pipelining is successful, it will either result in double buffering or fully unrolling the inner loops. Both options are usually preferred compared to other optimizations on the inner loop. 

Since getting the true value of design's objectives are time-consuming, building the dataset is the main bottleneck of our approach. After building an initial database as explained in Section~\ref{sec:db}, we exploit our DSE to augment the database. Note that the DSE wants to run the model on many of the unseen data points. It can perform well only when there are good representatives of all of the design choices in our database. On the other hand, if our DSE believes that an unseen design point is good even though in reality it is bad, it means that the model does not have a sufficient set of data to generalize for the whole space. Since these data points are the ones that made the model mispredict the results, they are more likely to build a better representation of data in the next round.

\section{Evaluation} \label{sec:eval}

\begin{table*}[!htb]
\footnotesize
\centering
\caption{Design space of the 9 kernels used for training our model}
\label{tbl:db}
\begin{tabular}{|c||c|c|c|c|c|c|c|c|c||c|}
\hline
\textbf{Kernel name} & aes & atax & gemm-blocked & gemm-ncubed & mvt & spmv-crs & spmv-ellpack & stencil & nw & Total\\ \hline
\textbf{\# pragmas} & 3 & 5 & 9 & 7 & 8 & 3 & 3 & 7 & 6 & - \\ \hline
\textbf{\# Designs} & 45 & 3,354 & 2,314 & 7,792 & 3,059,001 & 114 & 114 & 7,591 & 15,288 & 3,095,613\\ \hline
\makecell{\textbf{Initial database} \\ \textbf{(\# Total / \# Valid)}} & 15 / 15 & 605 / 101 & 616 / 149 & 432 / 149 & 571 / 180 & 98 / 35 & 114 / 60 & 1,066 / 281 & 911 / 66 & 4,428 / 1,036 \\ \hline
\makecell{\textbf{Final database} \\ \textbf{(\# Total / \# Valid)}} & 44 / 44 & 636 / 129 & 667 / 183 & 476 / 193 & 621 / 224 & 114 / 51 & 114 / 60 & 1,098 / 291 & 982 / 103 & 4,752 / 1,278 \\ \hline
\end{tabular}
\end{table*}


\subsection{Experimental Setup} \label{sec:eval_setup}
We choose our target kernels from the commonly-used Machsuite benchmark~\cite{machsuite}, and the Polyhedral benchmark suite (Polybench)~\cite{polybench}. To generate the initial database for each of the kernels, we extend the AutoDSE~\cite{sohrabizadeh2020autodse} framework as explained in Section~\ref{sec:db} with the Xilinx Virtex Ultrascale+ VCU1525 as the target FPGA. Our database consists of 9 kernels with different computation intensities including matrix and vector operations, stencil operation, encryption, and a dynamic programming application (\texttt{nw}). We wish to train a model to predict the \texttt{latency} in the form of \textit{cycle counts}, and the resource utilization for \texttt{DSP}, \texttt{BRAM}, \texttt{LUT}, and \texttt{FF}. Our framework is deployed and trained using PyTorch~\cite{pytorch}. 

Table~\ref{tbl:db} summarizes the number of pragmas, the total solution space size, the total number of configurations, and the number of valid configurations for each of our target kernels. As we explained in Section~\ref{sec:dse}, after running the DSE, we evaluate the top designs generated by the model using the HLS tool and add their true objectives to the database so that the model can refine its predictions. The final row in Table~\ref{tbl:db} summarizes the number of designs after these additions. In our database, the \texttt{latency} is in the range of 660 to 12,531,777 cycles. \texttt{DSP} counts are in the range of 0 to 28,672. The number of \texttt{BRAM} blocks ranges from 0 to 7,464. The \texttt{LUT} usage ranges from 913 to 2,639,487 and the \texttt{FF} ranges from 0 to 3,831,357.

\subsection{Model Evaluation} \label{sec:eval_model}
\subsubsection{Pre-processing the Data: }
 We start our training by pre-processing our data to limit their ranges so that they can contribute to the loss equally. For this matter, we normalize the number of resources each design takes by dividing them by the available number of resources on the FPGA. We use the following formula when the objective is \texttt{latency}:
\begin{equation}
    \label{eq:latency}
    T_{latency} = \log_2 {\frac{Normalization Factor}{latency}}
\end{equation}
therefore, the model spends more time on reducing the loss for large values of $T_{latency}$ which corresponds to low latency values, i.e., the high-performance designs. The $\log_2$ factor is used to make the data distribution more even since because of the intrinsic features of this problem, the number of high-performance values are limited and the data is originally biased towards low-performance ones. Although using the $\log_2$ factor can amplify the losses in prediction when the data is transformed back to its original range, it does not impact our final task which is to find the best design configuration. Note that, when running the DSE, the relative performance numbers are sufficient for us.

\subsubsection{Grouping the Objectives:}
\begin{figure}[!htb]
	\centering
	\includegraphics[width=0.6\columnwidth]{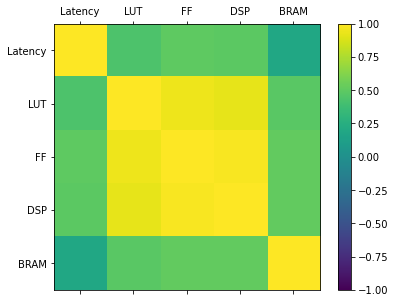} 
	\caption{Correlation matrix for our database}
	\label{fig:corr-matrix}
\end{figure}

Fig.~\ref{fig:corr-matrix} depicts the correlation matrix of the objectives in our database with 1 (-1) showing the highest (lowest) correlation. As the figure illustrates, the \texttt{LUT}, \texttt{FF}, and \texttt{DSP} values are highly correlated. The \texttt{latency} value has a weak correlation with each of the \texttt{LUT}, \texttt{FF}, and \texttt{DSP} values but it almost has no correlation with \texttt{BRAM}. As a result, the \texttt{latency}, \texttt{LUT}, \texttt{FF}, and \texttt{DSP} values can help each other to learn a better graph embedding as explained in Section~\ref{sec:MLP}. Consequently, we train two models to learn all the objectives, one is responsible solely to predict the \texttt{BRAM} utilization while the other one predicts the rest of the objectives.
\begin{table*}
  \begin{center}
    \caption{Model loss on our database. RMSE is used as the evaluation metric for the regression task. For the classification task, the percentage of misclassified cases are reported.
    }
    \label{tbl:regression}
    \begin{tabular}{|l|l||lllll|l||c|}
    \hline
    
    \textbf{Model} &
    \textbf{Method} &
    \textbf{Latency} &
    \textbf{DSP} &
    \textbf{LUT} &
    \textbf{FF} &
    \textbf{BRAM} & 
    \textbf{All} &
    \textbf{Misclassification}\\
      \hline \hline
    M1 & MLP-pragma         & 3.2756	& 0.5857	& 0.3115	& 0.2483 & 0.3356 & 4.7567 & 49.8\%\\
    M2 & MLP-pragma-program context      & 2.9444	& 0.4650	& 0.2401	& 0.1349 & 0.1597 & 3.9442 & 21.3\%\\
    \hline
  M3& \model - GCN      & 1.6825 	& 0.4265	& 0.1642	& 0.1277 & 0.1593 & 2.5602 & 22.9\%\\
    M4 & \model - GAT      & 1.1819 	& 0.2557	& 0.1266	& 0.1009 & 0.1178 & 1.7829 & 13.7\%\\
    M5 & \model - \tconv      & 1.1323	& 0.2540	& 0.1245	& 0.0938 & 0.1231 & 1.7277 & 10.6\%\\
    M6 & \model - \tconv + JKN      & 1.0846	& 0.2521 & 0.1112	& 0.0933 & 0.0912 & 1.6324 & 8.6\%\\
    \hline
    M7 & \model (\tconv + JKN + node att.)     & \textbf{0.5359}	& \textbf{0.1253}	& \textbf{0.0762}	& \textbf{0.0632} & \textbf{0.0515} & \textbf{0.8521} & \textbf{6.2\%}\\ \hline

    \end{tabular}
  \end{center}
\end{table*}

\subsubsection{Comparative Studies:}
To test whether our program representation is beneficial for our training task or not, we first test the performance of two models which only use an MLP network with no considerations for the graph structure. The first one (M1) just uses the pragma settings as the input to an MLP as in~\cite{kwon2020transfer}. The first layer of the MLP is to encode the pragmas of the design; hence, each application has a separate first layer. The rest of the network is shared among all the applications with a second level of MLP to predict the objectives. The second model (M2) takes all the nodes of the graph with their initial embeddings as the input but does not exploit the GNN techniques for updating the embeddings and rather only uses an MLP. It means that it does not see any of the node relations in the graph for transforming the features. Table~\ref{tbl:regression} summarizes the loss of these models and the breakdown of loss for each of the objectives for the regression task along with the percentage of misclassified cases for the classification task. The losses for the regression task are measured using root mean squared error (RMSE). As the results suggest, including the program context in the input is crucial for improving the accuracy of the model since it wants to predict the objectives across applications with different semantics.

Additionally, we assess the effect of our optimizations applied to the model. More specifically, we first tested our model's performance when it uses either of the GCN, GAT, or \tconv as the GNN layer with normal summation to create the graph-level embeddings (M3 to M5). As Table~\ref{tbl:regression} shows, since these models include the different flows (control, data, call, pragma) of the program using a graph structure in addition to the pragma/program context, they can decrease the loss. Besides, among the GNN models, the \tconv has the best performance due to the reasons mentioned in Section~\ref{sec:gnn-encoder}. We further evaluated the performance of our GNN model after adding each of the JKN (M6) and node attention (M7) layers. As the results in Table~\ref{tbl:regression} demonstrate, both of these optimizations are necessary since we are combining different types of nodes in our program representation. This validates our hypothesis that the model not only needs to have a way of deciding which instructions and pragmas are more important for the given application but also should be able to dynamically assign the ranges of neighborhood each node requires for updating its features.

\subsection{Results of Design Space Exploration} \label{sec:eval_dse}
Using our models, we are able to run 22 inferences per second. As a result, we can exhaustively search through all the design choices for our target kernel, except for \texttt{mvt}, in a few minutes. We set a time limit of one hour for running the DSE, which means we get to explore about 80K design choices. As a result, we adopt the heuristic we proposed in Section~\ref{sec:dse} to search through \texttt{mvt}.

\begin{figure}[!htb]
	\centering
	\includegraphics[width=0.95\columnwidth]{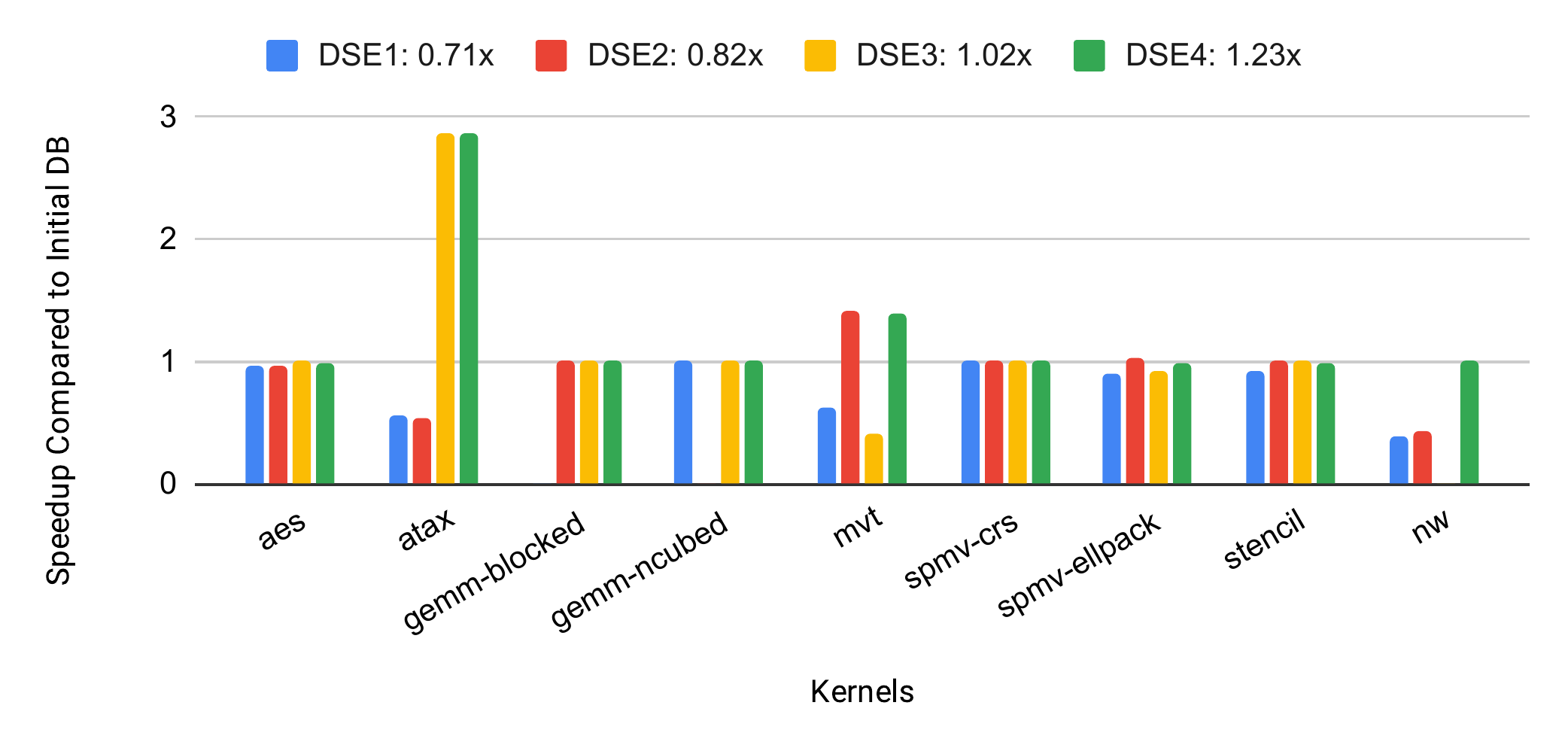} 
	\caption{The speedup of the GNN-DSE compared to the the best design in the initial database. After each round of DSE, the top designs are added to the database to refine the predictions.}
	\label{fig:dse_eval}
\end{figure}

As Table~\ref{tbl:db} suggests, our initial database has only $0.14\%$ of the total design choices which is a very small fraction. Hence, as we explained in Section~\ref{sec:dse}, we run the DSE on all the kernels and evaluate their top 10 designs using the HLS tool. Depending on how it performs, we add a various number of design points with their true objectives to the database.  Fig.~\ref{fig:dse_eval} depicts the speedup each kernel achieved compared to the best design in the initial database for different rounds of DSE. As the figure shows, after 3 rounds of database expansion (the last row of Table~\ref{tbl:db} summarizes the final number of designs in our database), the DSE can find a better or equal design configuration. The chart's legend summarizes the average speedup of all the kernels after each round.

\subsection{Results of Transfer Learning} \label{sec:eval_transfer_learning}
The final goal of our approach is to be extensible to the other kernels it has not seen. For testing this, we have chosen two new kernels from Polybench which were not included in our database: \texttt{2mm}, \texttt{bicg}, {\arxiv{and \texttt{gesummv}}}. The first one consists of two matrix multiplications, the second one is doing two matrix-vector multiplications, {\arxiv{and the third one consists of two matrix-vector multiplications and a weighted vector addition}}. Note that four of the kernels in our database are working with matrix-vector operations, although, in general, they have a different problem size and coding structure. 

Table~\ref{tbl:transfer_learning} summarizes the number of pragmas and the design configurations for each of these kernels. Since our models run in milliseconds order, the DSE is able to explore all the choices in only 2 {\arxiv{(1)}} minutes for \texttt{bicg} {\arxiv{\texttt{gesummv}}}. We adopt our heuristic explained in Section~\ref{sec:dse} for running the DSE on \texttt{2mm}, which has more than $492M$ design choices, for one hour. To measure the quality of top designs generated here, we ran the original explorer of AutoDSE for 21 hours, during which it explored {\arxiv{163/}}126/131 design points for {\arxiv{\texttt{gesummv}}}/\texttt{bicg}/\texttt{2mm} and achieved a speedup of {\arxiv{$18\times$}}/$26\times$/$350\times$ compared to the design with no optimizations. As the results in Table~\ref{tbl:transfer_learning} demonstrate, our approach achieved nearly the same performance for these kernels even though they were not included in the training set. Note that for the \texttt{bicg} kernel, it could achieve an even better performance since it was able to explore the whole space; while AutoDSE only explored $3.6\%$ of the space after running for 21 hours as it is dependent on invoking the time-consuming HLS tool.

\begin{table}[H]
\footnotesize
\centering
\caption{The performance of our approach on the target kernels that were not included in its database. The speedup numbers are with respect to a prior state-of-the-art work, AutoDSE~\cite{sohrabizadeh2020autodse}, after running it for 21 hours.}
\label{tbl:transfer_learning}
\begin{tabular}{|c||c|c|c|c|c|}
\hline
\textbf{Kernel} & \textbf{\# pragma} & \textbf{\# Design} & \makecell{\textbf{DSE Runtime} \\ \textbf{(min)} }&\textbf{\# Explored} & \textbf{Speedup} \\ \hline
\textbf{gesummv} & 4 & 1,581 & 1 &1,581 & $0.99\times$\\ \hline
\textbf{bicg} & 5 & 3,536 & 2 &3,536 & $1.05\times$\\ \hline
\textbf{2mm} & 14 & 492,787,501 & 60 & 78,676 & $0.98\times$\\ \hline
\end{tabular}
\end{table}
\section{Conclusion} \label{sec:conclusion}
In this work, we made our first step in modeling the HLS tool to assess a design's quality in milliseconds. For this matter, we developed a push-button framework called {\framework} to build a learning model for predicting the design's objectives. We proposed a graph-based program representation which includes both the program semantics and the candidate pragmas and implemented a GNN-based model to help us extract the required information for estimating our targets. We exploited our model to optimize the target applications by searching through their different design configurations and finding the Pareto-optimal ones. The experimental results show that {\framework} can build a model with high accuracy to be used among different domains. They also demonstrate that for each domain, we only need to include a few design configurations in the training set. Using them, {\framework} is able to not only find the Pareto-optimal designs quickly for the applications in its database, but also extend the knowledge it gained from them to optimize new applications from its existing domains. In the future, we will expand our tool to cover more domains.
\begin{acks}
We would like to thank Marci Baun for editing the paper. This work is supported by the CAPA award jointly funded by NSF (CCF-1723773) and Intel (36888881), the RTML award funded by NSF (CCF-1937599), and CDSC industrial partners\footnote{https://cdsc.ucla.edu/partners/}.
\end{acks}


\bibliographystyle{ACM-Reference-Format}
\bibliography{main}

\end{document}